\documentclass[journal,10.5pt]{IEEEtran}
\usepackage[T1]{fontenc}

\usepackage[pdftex]{graphicx}

\ifCLASSINFOpdf
\else
\fi
\usepackage{amsmath}
\usepackage{amsfonts}
\usepackage{amssymb}
\usepackage{bm}
\usepackage{color}
\usepackage{stfloats}
\usepackage{multicol}
\usepackage{CJK}
\usepackage{epstopdf}

\usepackage{epsf}
\usepackage{psfrag}
\usepackage{color}
\usepackage{graphicx,cite,epsfig,amssymb,amsmath,subfigure,url,stfloats,latexsym}
\usepackage{array}
\usepackage{pgfplots}
\usepackage{algorithm}
\usepackage[noend]{algpseudocode}
\usepackage{epstopdf}
\usepackage{amsfonts}
\usepackage{multirow}
\usepackage{mathrsfs}
\usepackage{subfigure}
\usepackage{algorithm}
\usepackage{algpseudocode}

\makeatother

\graphicspath{{./}{icons/}}
\graphicspath{{figures/}}
\usepackage[flushleft]{threeparttable}
\usepackage{verbatim}
\usepackage{lipsum}
\usepackage{fancyhdr}
\makeatletter
\def\headrule{\hrule \@height 0pt \@width \headwidth \vskip 3pt}
\makeatother

\begin{document}
%

\title{\LARGE Performance of OTFS-NOMA Scheme for Coordinated Direct and Relay Transmission Networks in High-Mobility Scenarios}

%
%

\author{Yao~Xu,
        Zhen~Du,
        Weijie~Yuan, ~\IEEEmembership{Member,~IEEE},
        Shaobo~Jia, ~\IEEEmembership{Member,~IEEE},\\
        and~Victor~C.~M.~Leung, ~\IEEEmembership{Life Fellow,~IEEE}
         \vspace*{-20pt}
\thanks{
Y.~Xu and Z.~Du are with  School of Electronic and Information Engineering,
Nanjing University of Information Science and Technology, Nanjing 210044,
China (email: yaoxu@nuist.edu.cn; duzhen@nuist.edu.cn).

Weijie Yuan is with the Department of Electronic and Electrical Engineering, Southern University of Science and Technology, Shenzhen
518055, China (email: yuanwj@sustech.edu.cn).

Shaobo Jia is with the School of Information Engineering, Zhengzhou University, Zhengzhou 450001, China (e-mail: ieshaobojia@
zzu.edu.cn).

V. C. M. Leung is with the College of Computer Science and Software
Engineering, Shenzhen University, Shenzhen 518060, China, and also with the
Department of Electrical and Computer Engineering, The University of British
Columbia, Vancouver, BC V6T 1Z4, Canada (e-mail: vleung@ieee.org).
}}
\maketitle

\thispagestyle{fancy}

%

\markboth{}%
{Shell \MakeLowercase{\textit{et al.}}: Bare Demo of IEEEtran.cls for IEEE Journals}
%



\begin{abstract}
In this paper, an orthogonal time frequency
space (OTFS) based non-orthogonal multiple access (NOMA) scheme is investigated for the coordinated direct and relay transmission system, where a source directly communicates with a near user with high mobile speed, and it needs the relaying assistance to serve the far user also having high mobility. Due to the coexistence of signal superposition coding and multi-domain transformation, the performance of OTFS-based NOMA is usually challenging to be measured from a theoretical perspective. To accurately evaluate the system performance of the proposed scheme, we derive the closed-form expressions for the outage probability and the outage sum rate by using the Inversion formula and characteristic function.
Numerical results verify the performance superiority and the effectiveness of the proposed scheme.
\end{abstract}


%
\IEEEpeerreviewmaketitle

\section{Introduction}
%
%
%
%






\IEEEPARstart{F}{uture} 6G wireless networks require superior spectral efficiency, larger coverage, and support for higher mobility \cite{R1}. Since coordinated direct and relay transmission (CDRT) using non-orthogonal multiple access (NOMA) can provide direct and relay communication links via power-domain superposition coding and successive interference cancellation (SIC), it is recognized as a potential enabling technology to simultaneously enhance the system spectral efficiency and coverage \cite{R2,R3}. NOMA scheme was first applied to the downlink two-user CDRT network \cite{R2}. Afterwards, NOMA-based CRDT was investigated in a variety of scenarios, including  bidirectional communication \cite{R4}, adaptive transmission \cite{R5}, cognitive network \cite{R6}, and physical layer security \cite{R7}.

Most existing NOMA-based CRDT schemes consider flat fading channels, which can provide valuable insights for low-mobility or stationary communication scenarios.
However, the doubly-selective fading channels in high-mobility scenarios may seriously affect the transmission reliability of the aforementioned schemes.
Fortunately, the recently proposed orthogonal time frequency space (OTFS) modulation can effectively combat the doubly-selective fading of high-mobility channels by utilizing relatively sparse and stable delay-Doppler domain signal processing \cite{R8,R9}. Therefore, some research efforts have been dedicated to studying OTFS-NOMA for direct transmission systems \cite{R10,R11}, but it usually uses simulations to analyze system performance because of the great challenge of accurate theoretical analysis caused by signal multi-domain transformation and superposition coding.

Motivated by these observations, we present a novel OTFS-NOMA scheme for a high-mobility CDRT system. To the best of our knowledge, NOMA-based CDRT has not yet been studied in high-mobility scenarios. The main contributions of this paper are three folds: 1) We propose an OTFS-NOMA scheme for a CDRT system with high-mobility, where a source employs the relaying and direct transmission modes to serve a far user and a near user, respectively, via OTFS transformation and power-domain superposition coding; 2) The closed-form expressions for the outage probability and outage sum rate of the proposed scheme are derived; 3) Numerical results are developed to illustrate that the proposed scheme can achieve superior outage performance for the far user and outage sum rate compered with the benchmarks.
\section{System Model}
As shown in Fig. 1, we consider a downlink NOMA-based CDRT system in high-mobility scenarios, where a source ($S$) directlys serve a near user ($U_c$) while communicating with a far user ($U_e$) via a decode-and-forward relaying node ($R$)  in two phases (i.e., $t_1$ and $t_2$ ). The wireless link between $S$ and $U_e$ is absent due to the significant obstruction or shadowing. Each node operates in half-duplex mode and employs a single antenna. For brevity, let subscripts $e$, $c$, $r$, and $s$ represent $U_e$, $U_c$, $R$, and $S$, respectively. When the transceiver moves at a relatively high speed, the multi-path effect and Doppler shift cause time dispersion and frequency dispersion, respectively, thus the wireless channels possess doubly selective fading characteristics in the time-frequency (TF) domain. The TF domain representation of the linear time-varying channel can be converted to a relatively sparse and stable representation in the delay-Doppler (DD) domain via the inverse symplectic fast Fourier transform (ISFFT). Assume that all the links experience doubly selective fading in the TF domain \cite{R9}, and the DD domain representation of the channel between $x\in\{s,r\}$ and $y\in\{ r,c,e\}$ in the $i$-th phase is given as
\begin{align}
h_{xy}^i(\tau ,v) = \sum\nolimits_{\omega  = 1}^{P_{xy}^i} {h_{xy}^{i,\omega }} \delta (\tau  - \tau _{xy}^{i,\omega })\delta (v - v_{xy}^{i,\omega })
\label{eq:1}
\end{align}
where $x\ne y$, $i \in \{ {t_1},{t_2}\}$, $\delta (\cdot)$ represents the Dirac delta function, $P_{xy}^i$ denotes the number of resolvable propagation paths, and $h_{xy}^{i,\omega}$, $\tau _{xy}^{i,\omega }$, and $v_{xy}^{i,\omega }$  denote the complex channel gain, the delay, and the Doppler shift corresponding to the $w$-th path, respectively. Particularly, the channel gain contains the path loss of the path component and satisfies $\sum\nolimits_{\omega  = 1}^{P_{xy}^i} {\{ |h_{xy}^{i,\omega }{|^2}\} }  = \varOmega _{xy}^i$. Without loss of generality, we assume that $\varOmega _{sr}^{{t_1}} < \varOmega _{sc}^{{t_1}}$, which may not require $\sum\nolimits_{\omega  = 1}^{P_{sr}^{{t_1}}} {|h_{sr}^{{t_1},\omega }{|^2}}< \sum\nolimits_{\omega  = 1}^{P_{sc}^{{t_1}}} {|h_{sc}^{{t_1},\omega }{|^2}}$.
\begin{figure*}[!t]
\centering
\includegraphics[width=4.5 in]{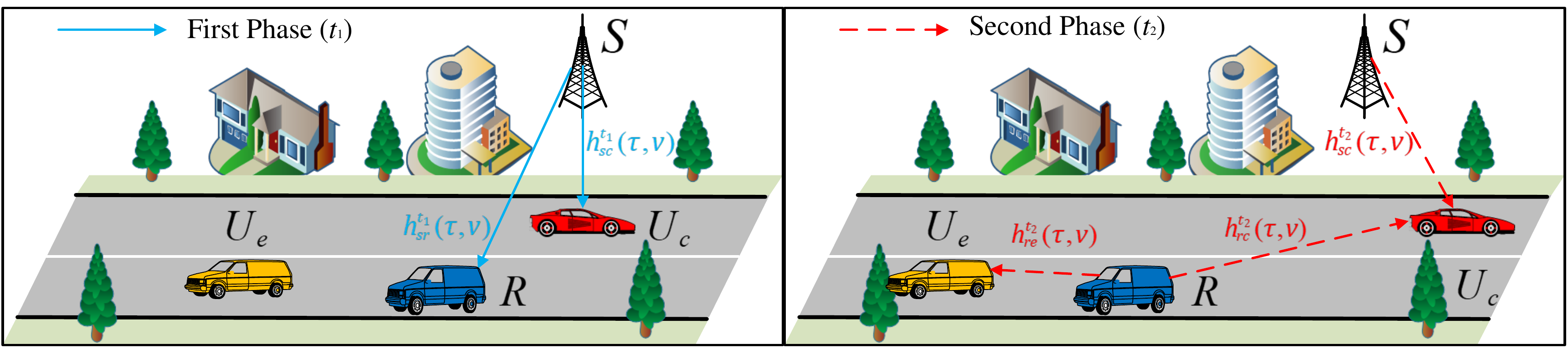}
\caption{System model. An illustration of the proposed OTFS-NOMA scheme for the CDRT system with one source, one relay, and two users (i.e., the far user and near user), where the multipath channels are indicated by arrows for simplicity.}
\label{Table1 model}
\vspace*{-15pt}
\end{figure*}
\vspace*{-5pt}
\section{OTFS-NOMA Transmission Scheme}
In this section, we design a novel OTFS-NOMA scheme for the CDRT system in high-mobility scenarios to ensure transmission reliability and spectrum efficiency. The parameter configuration and scheme design are detailed below.
\vspace*{-15pt}
\subsection{Parameter Configuration}
By using OTFS, the transmitter converts the modulated symbols located in the DD plane to the TF plane and then to the time domain by sequentially performing ISFFT and Heisenberg transform (i.e., OTFS modulation), and finally broadcasts the time domain signal. Correspondingly, the receiver obtains the DD domain signal from the time domain received signal via Wigner transform and symplectic fast Fourier transform (i.e., OTFS demodulation).

Consider a discrete TF plane with $M$ frequency subcarriers and $N$ time slots, where the subcarrier bandwidth and time slot duration are $\Delta f$ and $T = 1/\Delta f$, respectively. Therefore, the TF domain OTFS frame takes the total symbol duration of $NT$ and bandwidth $M\Delta f$, and the corresponding DD plane has the delay resolution of $1/(M\Delta f)$, Doppler resolution $1/(NT)$, and $MN$ DD bins. The parameters $T$ and $\Delta f$ satisfy $T\ge {\tau _{\max }}$ and $\Delta f \ge {v_{\max }}$, where ${\tau _{\max }} = \max \{ \tau _{xy}^{i,\omega },1 \le \omega  \le P_{xy}^i{\kern 1pt} {\kern 1pt} ,i \in {\bf{A}},x,y \in {\bf{B}},x \ne y\}$ and ${v_{\max }} = \max \{ v_{xy}^{i,\omega },1 \le \omega  \le P_{xy}^i,i \in {\bf{A}},x,y \in {\bf{B}},x \ne y\}$ denote the maximal delay spread and the the largest Doppler shift, respectively, ${\bf{A}}=\{ s,r,c,e\}$, and ${\bf{B}} = \{ {t_1},{t_2}\}$. Moreover, the delay $\tau _{xy}^{i,\omega}$ and the Doppler shift $v_{xy}^{i,\omega }$ in (1) can be written as $\tau _{xy}^{i,\omega } = (l_{xy}^{i,\omega } + \bar l_{xy}^{i,\omega })/(M\Delta f)$ and $v_{xy}^{i,\omega } = (k_{xy}^{i,\omega } + \bar k_{xy}^{i,\omega })/(NT)$, respectively, where the integers $k_{xy}^{i,\omega}$ and $l_{xy}^{i,\omega }$ represent the discrete Doppler and delay tap indices, respectively, and $\bar k_{xy}^{i,\omega }$ and $\bar l_{xy}^{i,\omega }$ denote the fractional Doppler shift and the fractional delay, respectively.
Since the practical broadband communication system can achieve high delay resolution, the fractional delay is usually ignored. Fractional Doppler can be effectively compensated by using the existing OTFS channel estimation or increasing the Doppler resolution \cite{R10, R12}. Therefore, as in \cite{R10}, we assume that $\bar l_{xy}^{i,\omega }=0$ and $\bar k_{xy}^{i,\omega }=0$.
\vspace*{-15pt}
\subsection{Scheme Design}
The proposed OTFS-NOMA scheme consists of two phases (i.e., $t_1$ and $t_2$), as detailed below.
\subsubsection{\bf{First Phase ($t_1$)}}
In $t_1$, the source $S$ places $MN$ superimposed symbols ${x_s}[k,l] = \sqrt {{\alpha _c}} {x_c}[k,l] + \sqrt {{\alpha _e}} {x_e}[k,l]$, $0 \le k \le N-1$, $0\le l \le M - 1$ on the DD grid, where ${x_c}[k,l]$ and ${x_e}[k,l]$ are the modulated symbols required by $U_c$ and $U_e$, respectively, and $\alpha_c$ and $\alpha_e$   represent the power allocation coefficients of ${x_c}[k,l]$ and ${x_e}[k,l]$, respectively. Since the far user $U_e$ needs the relaying assistance and the condition $\varOmega _{sr}^{{t_1}}<\varOmega_{sc}^{{t_1}}$ holds, the power allocation coefficients should meet ${\alpha_c}<{\alpha_e}$ and ${\alpha_c}+{\alpha_e}=1$ via downlink NOMA. By using the ISFFT, the DD domain symbols ${x_s}[k,l]$ can be converted to $MN$ TF domain symbols
\begin{align}
{X_s}[n,m] =& \frac{{\sqrt {{P_s}} }}{{NM}}\sum\nolimits_{k = 0}^{N - 1} {\sum\nolimits_{l = 0}^{M - 1} {(\sqrt {{\alpha _c}} {x_c}[k,l] + \sqrt {{\alpha _e}} {x_e}[k,l])} }\nonumber\\
 &\times \exp \Big\{ j2\pi \Big(\frac{{nk}}{N} - \frac{{ml}}{M}\Big)\Big\}
\label{eq:2}
\end{align}
 where $0 \le n \le N - 1$ and $0 \le m \le M - 1$, and $P_s$ denotes the transmit power of $S$.
After shaping the above TF domain symbols using the transmit pulse $g_{tx}(t)$ and performing the Heisenberg transform, the source $S$ can obtain a time domain signal and then broadcast the signal to $R$ and $U_c$.

In the receiver, the received time domain signal is shaped utilizing the receive pulse $g_{rx}(t)$, followed by Wigner transform. Similar to \cite{R10}, we assume that the orthogonality between the transmit and receive pulses is perfect. Therefore, the received TF signal at $R$ and $U_c$ in $t_1$ can be expressed as
\begin{align}
Y_{\bar x}^{{t_1}}[n,m] = H_{s\bar x}^{{t_1}}[n,m]{X_s}[n,m] + W_{\bar x}^{{t_1}}[n,m]
\label{eq:3}
\end{align}
where $\bar x \in \{ r,c\}$, $W_{\bar x}^{{t_1}}[n,m]$ represents the TF domain additive white Gaussian noise at the receiver $\bar x$ in the first phase, and $H_{s\bar x}^{{t_1}}[n,m] = \int {\int {h_{s\bar x}^{{t_1}}(\tau ,v)} } \exp (j2\pi vnT) \exp ( - j2\pi (v + m\Delta f)\tau )d\tau dv$.
After the TF domain signal $Y_{\bar x}^{{t_1}}[n,m]$ is processed by using the SFFT, the receiver $\bar x$ can obtain the DD domain signal as
\begin{align}
\!\!y_{\bar x}^{{t_1}}[k,l] \!\!=&\!\!\sum\nolimits_{n = 0}^{N - 1} \sum\nolimits_{m = 0}^{M - 1} Y_{\bar x}^{{t_1}}[n,m]\\ \nonumber
&\times\exp \Big\{ \!\!\!-\!j2\pi \Big(\!\frac{{nk}}{N}\!\!-\!\frac{{ml}}{M}\Big)\!\Big\}
\label{eq:4}
\end{align}

Substituting (1)-(3) into (4), $y_{\bar x}^{{t_1}}[k,l]$ can be rewritten as
\begin{align}
y_{\bar x}^{{t_1}}[k,l] \!= &\sqrt {{P_s}} \sum\nolimits_{\omega  = 1}^{P_{s\bar x}^{{t_1}}} {\tilde h_{s\bar x}^{{t_1},\omega }}\\ \nonumber
&\times \big\{ \sqrt {{\alpha _c}} {x_c}\big[{\big(k - k_{s\bar x}^{{t_1},\omega }\big)_N},{\big(l - l_{s\bar x}^{{t_1},\omega }\big)_M}\big]\\ \nonumber
& + \sqrt {{\alpha _e}} {x_e}\big[{\big(k - k_{s\bar x}^{{t_1},\omega }\big)_N},{\big(l - l_{s\bar x}^{{t_1},\omega }\big)_M}\big]\big\} + w_{\bar x}^{{t_1}}[k,l]
\label{eq:5}
\end{align}
where $\tilde h_{s\bar x}^{{t_1},\omega } = h_{s\bar x}^{{t_1},\omega }\exp ( - j2\pi \tau _{s\bar x}^{{t_1},\omega }v_{s\bar x}^{{t_1},\omega })$, ${(\cdot)_M}$ is the modulo $M$ operator, and $w_{\bar x}^{{t_1}}[k,l]$ denotes the complex Gaussian noise having zero mean and variance ${\sigma ^2}$. Note that the distribution of $\tilde h_{s\bar x}^{{t_1},\omega }$ satisfies $\tilde h_{s\bar x}^{{t_1},\omega } \sim {\cal {CN}}(0,\varOmega _{xy}^{i,\omega })$ due to $h_{xy}^{i,\omega } \sim {\cal {CN}}(0,\varOmega _{xy}^{i,\omega })$. To facilitate analysis, the vector form of $y_{\bar x}^{{t_1}}[k,l]$ can be expressed as
\begin{align}
{\bf{y}}_{\bar x}^{{t_1}} = {\bf{H}}_{s\bar x}^{{t_1}}(\sqrt {{\alpha _c}{P_s}} {{\bf{x}}_c} + \sqrt {{\alpha _e}{P_s}} {{\bf{x}}_e}) + {\bf{w}}_{\bar x}^{{t_1}}
\end{align}
where ${\mathbf{y}}_{\bar x}^{{t_1}} \in {\mathbb{C}^{NM \times 1}}$, ${{\mathbf{x}}_c} \in {\mathbb{C}^{NM \times 1}}$, ${{\mathbf{x}}_e} \in {\mathbb{C}^{NM \times 1}}$, ${\mathbf{w}}_{\bar x}^{{t_1}} \in {\mathbb{C}^{NM \times 1}}$, and ${\mathbf{H}}_{s\bar x}^{{t_1}} \in {\mathbb{C}^{NM \times NM}}$ is the effective DD domain channel matrix. The $(lN + k)$-th element of ${\mathbf{y}}_{\bar x}^{{t_1}}$ is $y_{\bar x}^{{t_1}}[k,l]$ for $0 \leqslant k \leqslant N-1$ and $0 \leqslant l \leqslant M-1$. Similarly, the vectors ${{\mathbf{x}}_c}$, ${{\mathbf{x}}_e}$, and ${\mathbf{w}}_{\bar x}^{{t_1}}$ can be constructed from ${x_c}[k,l]$, ${x_e}[k,l]$, and $w_{\bar x}^{{t_1}}[k,l]$, respectively. The complex Gaussian noise vector ${\mathbf{w}}_{\bar x}^{{t_1}}$ has the independent and identically distributed (i.i.d.) element with zero mean and variance ${\sigma ^2}$, and the elements of ${{\mathbf{x}}_c}$ and ${{\bf{x}}_e}$ are normalized with i.i.d. distribution.
The zero-forcing (ZF) equalizer is commonly used in OFDM or OTFS systems due to low complexity \cite{R13}. Similarly, the ZF algorithm is employed to equalize the received signals in this paper. After ${\bf{y}}_{\bar x}^{{t_1}}$ is multiplied by the ZF coefficient ${\bf{E}}_{s\bar x}^{{t_1}} = {({\bf{H}}{_{s\bar x}^{{{t_1}}H}}{\bf{H}}_{s\bar x}^{{t_1}})^{ - 1}}{\bf{H}}{_{s\bar x}^{{{t_1}}H}}$, the equalized signal ${\bf{\tilde y}}_{\bar x}^{{t_1}}$ can be written as
\begin{align}
{\bf{\tilde y}}_{\bar x}^{{t_1}} = \sqrt {{\alpha _c}{P_s}} {{\bf{x}}_c} + \sqrt {{\alpha _e}{P_s}} {{\bf{x}}_e} + {\bf{\tilde w}}_{\bar x}^{{t_1}}
\label{eq:7}
\end{align}
where ${\bf{\tilde w}}_{\bar x}^{{t_1}} = {\bf{E}}_{s\bar x}^{{t_1}}{\bf{w}}_{\bar x}^{{t_1}}$ is the equivalent noise having the covariance matrix of ${{\bf{C}}_{{\bf{\tilde w}}_{\bar x}^{{t_1}}}} = E[{\bf{\tilde w}}_{\bar x}^{{t_1}}{\bf{\tilde w}}{_{\bar x}^{{{t_1}} H}}] = {\sigma ^2}{({\bf{H}}{_{s\bar x}^{{{t_1}} H}}{\bf{H}}_{s\bar x}^{{t_1}})^{-1}}$.
Note that the channel matrix ${\bf{H}}_{s\bar x}^{{t_1}}$ has a doubly-block circulant structure, which means that ${\bf{H}}_{s\bar x}^{{t_1}}$ has $M$ circulant blocks and each block is a  circulant matrix with the size of $N \times N$. The matrix ${\bf{H}}_{s\bar x}^{{t_1}}$ can be decomposed into ${\bf{H}}_{s\bar x}^{{t_1}} = {{\bf{\Lambda }}^H}{\bf{\Xi }}_{s\bar x}^{{t_1}}{\bf{\Lambda }}$, and thus we have ${\bf{H}}{_{s\bar x}^{{t_1}H}}{\bf{H}}_{s\bar x}^{{t_1}} = {{\bf{\Lambda }}^H}{\bf{\Xi }}{_{s\bar x}^{{t_1}H}}{\bf{\Xi }}_{s\bar x}^{{t_1}}{\bf{\Lambda }}$, where ${\mathbf{\Lambda }} = {{\mathbf{F}}_M} \otimes {{\mathbf{F}}_N} \in {\mathbb{C}^{NM \times NM}}$, ${\bf{\Xi }}_{s\bar x}^{{t_1}} = diag[\lambda {_{s\bar x,1}^{{t_1}}},\lambda {_{s\bar x,2}^{{t_1}}}, \cdots ,\lambda {_{s\bar x,{NM}}^{{t_1}}}]$, and $\lambda _{s\bar x,w}^{{t_1}}$, $1 \le w \le NM$ is the eigenvalue of ${\bf{H}}_{s\bar x}^{{t_1}}$. Besides, the entries of ${\bf{\Xi }}{_{s\bar x}^{{t_1}H}}{\bf{\Xi }}_{s\bar x}^{{t_1}}$ are the eigenvalues of ${\bf{H}}{_{s\bar x}^{{t_1}H}}{\bf{H}}_{s\bar x}^{{t_1}}$, i.e., $|\lambda _{s\bar x,w}^{{t_1}}{|^2},1 \le w \le NM$ \cite{R13}.

According to downlink NOMA, $U_c$ performs SIC to decode ${{\bf{x}}_c}$ after removing the interference ${{\bf{x}}_e}$, while $R$ directly decode ${{\bf{x}}_e}$ by treating ${{\bf{x}}_c}$ as noise.
Therefore, based on (7) and ${{\bf{C}}_{{\bf{\tilde w}}_{\bar x}^{{t_1}}}}$, the signal-to-interference-plus-noise ratios (SINRs) for the receiver $\bar x \in \{ r,c\}$ to decode the $k$-th symbols of ${{\bf{x}}_c}$ and ${{\bf{x}}_e}$ (i.e., ${{\bf{x}}_{c,k}}$ and ${{\bf{x}}_{e,k}}$) in $t_1$ can be written as
\begin{align}
\gamma _{\bar x,{{\bf{x}}_{e,k}}}^{{t_1}} &= \frac{{{\alpha _e}{\rho _s}}}{{{\alpha _c}{\rho _s} + {{[{{({\bf{H}}{{_{s\bar x}^{{t_1}H}}}{\bf{H}}_{s\bar x}^{{t_1}})}^{ - 1}}]}_{k,k}}}}\\ \nonumber
 &= \frac{{{\alpha _e}{\rho _s}}}{{{\alpha _c}{\rho _s} + {{[{{({{\bf{\Lambda }}^H}{\bf{\Xi }}{{_{s\bar x}^{{t_1}H}}}{\bf{\Xi }}_{s\bar x}^{{t_1}}{\bf{\Lambda }})}^{ - 1}}]}_{k,k}}}} = \frac{{{\alpha _e}{\rho _s}}}{{{\alpha _c}{\rho _s} + \frac{{\varTheta _{sc}^{{t_1}}}}{{MN}}}}
\\
\gamma _{c,{{\bf{x}}_{c,k}}}^{{t_1}} &= \frac{{{\alpha _c}{\rho _s}}}{{{{[{{({\bf{H}}{{_{sc}^{{t_1}H}}}{\bf{H}}_{sc}^{{t_1}})}^{ - 1}}]}_{k,k}}}}
= \frac{{MN{\alpha _c}{\rho _s}}}{{\varTheta _{sc}^{{t_1}}}}
\label{eq:8,9}
\end{align}
where $\varTheta _{s\bar x}^{{t_1}} = \sum\nolimits_{w = 1}^{MN} {|\lambda {{_{s\bar x,}^{{t_1}}}_w}{|^{-2}}}$, $\varTheta _{sc}^{{t_1}} = \sum\nolimits_{w = 1}^{MN} {|\lambda {{_{sc,}^{{t_1}}}_w}{|^{-2}}}$, and ${\rho_s} = \frac{{{P_s}}}{{{\sigma ^2}}}$ is the transmit signal-to-noise ratio (SNR).
\subsubsection{\bf{Second Phase ($t_2$)}}
During the second phase, the relay $R$ attempts to forward the decoded signal ${{\bf{x}}_e}$ via OTFS. Specifically, $MN$ decoded symbols ${x_e}[k,l]$, $0 \le k \le N - 1$, $0 \le l \le M - 1$  are placed on the DD grid, and then $R$ uses OTFS modulation to transmit the corresponding time domain signal. Meanwhile, to improve the spectral efficiency, the source $S$ broadcasts $MN$ new DD domain symbols ${\bar x_c}[k,l]$, $0 \le k \le N - 1$, $0 \le l \le M - 1$ by employing OTFS modulation, where ${\bar x_c}[k,l]$ are the desired signals for $U_c$ in $t_2$. Similarly to (4)-(6), $U_c$ and $R$ can adopt OTFS demodulation to obtain the received DD domain signal in vector form as
\begin{align}
{\bf{y}}_c^{{t_2}}&= {\bf{H}}_{sc}^{{t_2}}\sqrt {{P_s}} {{\bf{\bar x}}_c} + {\bf{H}}_{rc}^{{t_2}}\sqrt {{P_r}} {{\bf{x}}_e} + {\bf{w}}_c^{{t_2}}\\
{\bf{y}}_e^{{t_2}}&= {\bf{H}}_{re}^{{t_2}}\sqrt {{P_r}} {{\bf{x}}_e} + {\bf{w}}_e^{{t_2}}
\label{eq:10,11}
\end{align}
where ${\mathbf{y}}_c^{{t_2}} \in {\mathbb{C}^{NM \times 1}}$, ${\mathbf{y}}_e^{{t_2}} \in {\mathbb{C}^{NM \times 1}}$, and $P_r$ is the transmit power of $R$. The vectors ${\mathbf{w}}_c^{{t_2}} \in {\mathbb{C}^{NM \times 1}}$ and ${\mathbf{w}}_e^{{t_2}} \in {\mathbb{C}^{NM \times 1}}$  are i.i.d. complex Gaussian noise having zero mean and  variance ${\sigma ^2}$, and ${\mathbf{H}}_{sc}^{{t_2}} \in {\mathbb{C}^{NM \times NM}}$, ${\mathbf{H}}_{rc}^{{t_2}} \in {\mathbb{C}^{NM \times NM}}$ and ${\mathbf{H}}_{re}^{{t_2}} \in {\mathbb{C}^{NM \times NM}}$ denote the corresponding effective DD domain channel matrices. The $(lN + k)$-th normalized element of ${{\mathbf{\bar x}}_c} \in {\mathbb{C}^{NM \times 1}}$ is ${\bar x_c}[k,l]$ for $0 \leqslant k \leqslant N - 1$ and $0 \leqslant l \leqslant M - 1$, and the vector ${{\mathbf{\bar x}}_e} \in {\mathbb{C}^{NM \times 1}}$ has a similar structure to ${{\mathbf{\bar x}}_c}$.

After estimating ${\bf{H}}_{rc}^{{t_2}}\sqrt {{P_r}}$, $U_c$ can utilize the obtained signal ${{\bf{x}}_e}$ in $t_1$ to remove the interference term ${\bf{H}}_{rc}^{{t_2}}\sqrt {{P_r}} {{\bf{x}}_e}$ in (10), and ${\bf{y}}_c^{{t_2}}$ becomes ${\bf{y}}_c^{{t_2}} = {\bf{H}}_{sc}^{{t_2}}\sqrt {{P_s}} {{\bf{\bar x}}_c} + {\bf{w}}_c^{{t_2}}$. Moreover, $U_c$ and $U_e$ use ZF coefficients ${\bf{E}}_{sc}^{{t_2}} = {({\bf{H}}{_{sc}^{{t_2}H}}{\bf{H}}_{sc}^{{t_2}})^{ - 1}}{\bf{H}}{_{sc}^{{t_2}H}}$ and ${\bf{E}}_{re}^{{t_2}} = {({\bf{H}}{_{re}^{{t_2}H}}{\bf{H}}_{re}^{{t_2}})^{ - 1}}{\bf{H}}{_{re}^{{t_2}H}}$ to equalize ${\bf{y}}_c^{{t_2}}$ and ${\bf{y}}_e^{{t_2}}$, respectively. The corresponding equalized signal ${\bf{\tilde y}}_c^{{t_2}}$ and ${\bf{y}}_e^{{t_2}}$ can be written as ${\bf{\tilde y}}_c^{{t_2}} = \sqrt {{P_s}} {{\bf{\bar x}}_c} + {\bf{\tilde w}}_c^{{t_2}}$ and ${\bf{\tilde y}}_e^{{t_2}} = \sqrt {{P_r}} {{\bf{x}}_e} + {\bf{\tilde w}}_e^{{t_2}}$, respectively, where the equivalent noise vectors ${\bf{\tilde w}}_c^{{t_2}} = {\bf{E}}_{sc}^{{t_2}}{\bf{w}}_c^{{t_2}}$ and ${\bf{\tilde w}}_e^{{t_2}} = {\bf{E}}_{re}^{{t_2}}{\bf{w}}_e^{{t_2}}$ have the covariance matrices of ${{\bf{C}}_{{\bf{\tilde w}}_c^{{t_2}}}} = {\sigma ^2}{({\bf{H}}{_{sc}^{{t_2}H}}{\bf{H}}_{sc}^{{t_2}})^{ - 1}}$ and ${{\bf{C}}_{{\bf{\tilde w}}_e^{{t_2}}}} = {\sigma ^2}{({\bf{H}}{_{re}^{{t_2}H}}{\bf{H}}_{re}^{{t_2}})^{ - 1}}$, respectively. Similarly, the SINR for $U_c$ and $U_e$ to decode the $k$-th symbols of  ${{\bf{\bar x}}_c}$ and ${{\bf{x}}_e}$ (i.e., ${{\bf{\bar x}}_{c,k}}$ and ${{\bf{x}}_{e,k}}$) in $t_2$ can be expressed as
\begin{align}
\gamma _{c,{{{\bf{\bar x}}}_{c,k}}}^{{t_2}}&= \frac{{{\rho _s}}}{{{{\big[{{\big({\bf{H}}{{_{sc}^{{t_2}H}}}{\bf{H}}_{sc}^{{t_2}}\big)}^{ - 1}}\big]}_{k,k}}}} = \frac{{{\rho _s}MN}}{{\varTheta _{sc}^{{t_2}}}}\\
\gamma _{e,{{\bf{x}}_{e,k}}}^{{t_2}}&= \frac{{{\rho _r}}}{{{{\big[{{\big({\bf{H}}{{_{re}^{{t_2}H}}}{\bf{H}}_{re}^{{t_2}}\big)}^{ - 1}}\big]}_{k,k}}}} = \frac{{{\rho _r}MN}}{{\varTheta _{re}^{{t_2}}}}
\label{eq:12,13}
\end{align}
where $\varTheta _{sc}^{{t_2}} = \sum\nolimits_{w = 1}^{MN} {|\lambda {{_{sc,}^{{t_2}}}_w}{|^{{\rm{ - }}2}}}$, $\varTheta _{re}^{{t_2}} = \sum\nolimits_{w = 1}^{MN} {|\lambda {{_{re,}^{{t_2}}}_w}{|^{ - 2}}} $, and $\lambda _{sc,w}^{{t_2}}$ and $\lambda _{re,w}^{{t_2}}$, $1 \le w \le NM$, are the eigenvalues for ${\bf{H}}_{sc}^{{t_2}}$ and ${\bf{H}}_{re}^{{t_2}}$, respectively.
\section{Performance Analysis}
This section calculates the closed-form expressions of the outage probability and outage sum rate for the proposed OTFS-NOMA scheme in high-mobility scenarios.

Based on (8), (9), (12), (13), and \cite{R13}, the variable $\varTheta _{\dot x}^i$, $i \in \{ {t_1},{t_2}\}$, $\dot x \in \{ sc,sr,re\}$ is a linear combination of $NM$ inverse gamma distribution variables $|\lambda {_{\dot x,w}^i}{|^{-2}}$, $1 \le w \le NM$, having $G_{\dot x}^i$ distinct values, and $|\lambda {_{\dot x,w}^i}{|^{-2}} \sim {\cal {IG}}(1,\varOmega _{\dot x}^i)$ and $|\lambda {_{\dot x,w}^i}| \sim {\cal {E}} (\frac{1}{{\varOmega _{\dot x}^i}})$, where ${\cal {IG}}(1,\varOmega _{\dot x}^i)$ denotes the inverse gamma distribution with a scale parameter $\varOmega _{\dot x}^i$ and a unity shape parameter, and ${\cal {E}} (\frac{1}{{\varOmega _{\dot x}^i}})$ is the exponential distribution with a rate parameter $\frac{1}{{\varOmega_{\dot x}^i}}$.

 Particularly, the value of $G_{\dot x}^i$ satisfying $1 \le G_{\dot x}^i \le NM$ is determined by the structure characteristics of the DD domain channel matrix ${\bf{H}}_{\dot x}^i$. Therefore, the variable $\varTheta_{\dot x}^i$ can be rewritten as $\varTheta _{\dot x}^i = \sum\nolimits_{\dot w = 1}^{G_{\dot x}^i} {{C_{G_{\dot x}^i,\dot w}}|\lambda {{_{\dot x,{\dot w}}^i}}{|^{ - 2}}}$, where ${C_{G_{\dot x}^i,\dot w}}$, $1 \le \dot w \le G_{\dot x}^i$, denotes the number of the $\dot w$-th distinct $|\lambda {_{sc,{\dot w}}^{{t_1}}}{|^{-2}}$ and meets the condition $\sum\nolimits_{\dot w = 1}^{G_{\dot x}^i} {{C_{G_{\dot x}^i,\dot w}}}=NM$.

Using [14, eqs. (5) and (12)], we can obtain the characteristic function for $\varTheta _{\dot x}^i$ as
\begin{align}
\bar \psi _{\dot x}^i(t) = {\prod}_{\dot w = 1}^{G_{\dot x}^i}\psi \big\{ {C_{G_{\dot x}^i,\dot w}}t,\varOmega _{\dot x}^i\big\}
\label{eq:14}
\end{align}
where the function $\psi \{ t,\varOmega \}$ is defined as $\psi \{ t,\varOmega \}  \buildrel \Delta \over = \frac{{2\sqrt { - jt\varOmega } {K_1}(2\sqrt {-jt\varOmega} /\varOmega )}}{\varOmega }$, and ${K_1}( \cdot )$ is the modified Bessel function of second kind.
To facilitate the analysis, we define the following function as
\begin{align}
&\Phi (z,\mu ,I,\bar \psi (t)) = 0.5 - \frac{1}{\pi }\\ \nonumber
&\times{\sum\limits}_{i = 0}^I {{\cal I}{\cal M}} \Big(\frac{{\exp ( - j(i + 0.5)\mu z)\bar \psi \{ (i + 0.5)\mu \} \} }}{{i + 0.5}}\Big)
\label{eq:15}
\end{align}
where ${\cal I}{\cal M}( \cdot )$ represents the operation of taking imaginary part.
By using the above analysis, the performance for the outage probability and outage sum rate are derived in general case (i.e., $ 1\le G_{\dot x}^i \le NM$) and special case (i.e., ${G_{\dot x}^i}=1$) respectively, detailed as below.
\subsection{Outage Probability}
\subsubsection{\bf{General Case}}
Based on (8) and (9), both the conditions $\frac{1}{2}\log (1 + \gamma _{c,{{\bf{x}}_{e,k}}}^{{t_1}}) > {R_{{{\bf{x}}_{e,k}}}}$ and $\frac{1}{2}\log (1 + \gamma _{c,{{\bf{x}}_{c,k}}}^{{t_1}}) > {R_{{{\bf{x}}_{c,k}}}}$ should hold if the receiver $U_c$ can decode ${{\bf{x}}_{c,k}}$, $1 \le k \le NM$ successfully via SIC in $t_1$,  where ${R_{{{\bf{x}}_{e,k}}}}$ and ${R_{{{\bf{x}}_{c,k}}}}$ denote the target date rates for ${{\bf{x}}_{c,k}}$ and ${{\bf{x}}_{e,k}}$, respectively.
Therefore, the outage probability for $U_c$ to decode ${{\bf{x}}_{c,k}}$ in $t_1$ is written as
\begin{align}
P_{{{\bf{x}}_{c,k}}}^{out} &= 1 - \Pr (\gamma _{c,{{\bf{x}}_{e,k}}}^{{t_1}} > {\varphi _{{{\bf{x}}_{e,k}}}},\gamma _{c,{{\bf{x}}_{c,k}}}^{{t_1}} > {\varphi _{{{\bf{x}}_{c,k}}}}) \\ \nonumber
&= 1 - \Pr (\varTheta _{sc}^{{t_1}} < {\xi _1})
\label{eq:16}
\end{align}
where ${\xi _1} = \min \Big\{ \frac{{MN({\alpha _e} - {\varphi _{{{\bf{x}}_{e,k}}}}{\alpha _c}){\rho _s}}}{{{\varphi _{{{\bf{x}}_{e,k}}}}}},\frac{{MN{\alpha _c}{\rho _s}}}{{{\varphi _{{{\bf{x}}_{c,k}}}}}}\Big\}$,
${\varphi _{{{\bf{x}}_{e,k}}}} = {2^{2{R_{{{\bf{x}}_{e,k}}}}}}-1$, and ${\varphi _{{{\bf{x}}_{c,k}}}} = {2^{2{R_{{{\bf{x}}_{c,k}}}}}} - 1$.
By using (15) and [14, eq. (29)], the cumulative distribution function of $\varTheta _{sc}^{{t_1}}$ is obtained as $\Phi ({\xi _1},{\mu _1},{I_1},\bar \psi _{sc}^{{t_1}}(t))$, and thus $P_{{{\bf{x}}_{c,k}}}^{out}$ in (16) can be rewritten as
\begin{align}
P_{{{\bf{x}}_{c,k}}}^{out} = \left\{ {\begin{array}{*{20}{l}}
{1 - \Phi ({\xi _1},{\mu _1},{I_1},\bar \psi _{sc}^{{t_1}}(t)),}&{{\alpha _e}/{\alpha _c} > {\varphi _{{{\bf{x}}_{e,k}}}}}\\
{1,}&{{{\alpha _e}/{\alpha _c} \le {\varphi _{{{\bf{x}}_{e,k}}}}}}
\end{array}} \right.
\end{align}
where ${\mu_1}$ and ${I_1}$ denote the step parameter and complexity-accuracy tradeoff parameter, respectively. Note that the setting of power allocation coefficients should satisfy ${{\alpha _e}/{\alpha _c} > {\varphi _{{{\bf{x}}_{e,k}}}}}$ for successful NOMA implement.

According to the proposed scheme, the user $U_e$ tries to decode ${{\bf{x}}_{e,k}}$ in $t_2$ after $R$ can decode ${{\bf{x}}_{e,k}}$ successfully in $t_1$. In other words, the non-outage event for the information transmission of ${{\bf{x}}_{e,k}}$ occurs when the inequality conditions $\frac{1}{2}\log (1 + \gamma _{r,{{\bf{x}}_{e,k}}}^{{t_1}}) > {R_{{{\bf{x}}_{e,k}}}}$ and $\frac{1}{2}\log (1 + \gamma _{e,{{\bf{x}}_{e,k}}}^{{t_2}}) > {R_{{{\bf{x}}_{e,k}}}}$ are met.
Consequently, based on (15) and [14, eq. (29)], the outage probability for ${{\bf{x}}_{e,k}}$ at $U_e$ in $t_2$ can be given as
\begin{align}
P_{{{\bf{x}}_{e,k}}}^{out} &= 1 - \Pr (\gamma _{r,{{\bf{x}}_{e,k}}}^{{t_1}} > {\varphi _{{{\bf{x}}_{e,k}}}},\gamma _{e,{{\bf{x}}_{e,k}}}^{{t_2}} > {\varphi _{{{\bf{x}}_{e,k}}}})\\ \nonumber
 &= 1 - \Pr (\varTheta _{sr}^{{t_1}} < {\xi _2},\varTheta _{re}^{{t_2}} < {\xi _3})\\ \nonumber
 &= \left\{ {\begin{array}{*{20}{l}}
\begin{array}{l}
{\kern -5.5pt} 1- \Phi ({\xi _2},{\mu _2},{I_2},\bar \psi _{sr}^{{t_1}}(t))\\
{\kern -0.5pt} \times \Phi ({\xi _3},{\mu _3},{I_3},\bar \psi _{re}^{{t_2}}(t)),
\end{array}&{{\alpha _e}/{\alpha _c} > {\varphi _{{{\bf{x}}_{e,k}}}}}\\
{1,}&{{\alpha _e}/{\alpha _c} \le {\varphi _{{{\bf{x}}_{e,k}}}}}
\end{array}} \right.
\end{align}
where ${\xi _2} = \frac{{MN({\alpha _e} - {\alpha _c}{\varphi _{{{\bf{x}}_{e,k}}}}){\rho _s}}}{{{\varphi _{{{\bf{x}}_{e,k}}}}}}$, ${\xi _3} = \frac{{MN{\rho _r}}}{{{\varphi _{{{\bf{x}}_{e,k}}}}}}$, ${\rho _r}{\rm{ = }}\frac{{{P_r}}}{{{\sigma ^2}}}$, and ${\mu _2}$ and ${\mu _3}$ are the step parameters, and ${I_2}$ and ${I_3}$ denote the complexity-accuracy tradeoff parameters. The theoretical result in (18) shows that the power allocation coefficients should satisfy ${\alpha _e}/{\alpha _c} > {\varphi _{{{\bf{x}}_{e,k}}}}$ to implement NOMA in practice.

Moreover, $U_c$ cannot decode ${{\bf{\bar x}}_{c,k}}$ in $t_2$ when the condition $\frac{1}{2}\log (1 + \gamma _{c,{{{\bf{\bar x}}}_{c,k}}}^{{t_2}}) < {R_{{{{\bf{\bar x}}}_{c,k}}}}$ holds, where ${R_{{{{\bf{\bar x}}}_{c,k}}}}$ denotes the target date rate of ${{\bf{\bar x}}_{c,k}}$. Similarly, we can utilize (12) and (15) to obtain the outage probability for $U_c$ to decode ${{\bf{\bar x}}_{c,k}}$ as
\begin{align}
P_{{{{\bf{\bar x}}}_{c,k}}}^{out}& = \Pr (\gamma _{c,{{{\bf{\bar x}}}_{c,k}}}^{{t_2}} < {\varphi _{{{{\bf{\bar x}}}_{c,k}}}})\\ \nonumber
 &= \Pr (\Theta _{sc}^{{t_2}} > {\xi _4})= 1 - \Phi ({\xi _4},{\mu _4},{I_4},\bar \psi _{sc}^{{t_2}}(t))
\end{align}
where ${\xi _4} = \frac{{MN{\rho _s}}}{{{\varphi _{{{{\bf{\bar x}}}_{c,k}}}}}}$, ${\varphi _{{{{\bf{\bar x}}}_{c,k}}}} = {2^{2{R_{{{{\bf{\bar x}}}_{c,k}}}}}} - 1$, and ${\mu _4}$ and ${I_4}$ are the step parameter and complexity-accuracy tradeoff parameter, respectively.
\subsubsection{\bf{Special Case}}
 In this case, the variable $\varTheta _{\dot x}^i$ can be given as $\varTheta _{\dot x}^i = MN|\lambda _{\dot x}^i{|^{ - 2}}$, where $|\lambda _{\dot x}^i| \sim {\cal {E}} (\frac{1}{{\varOmega _{\dot x}^i}})$. Therefore, the SINRs in (8) and (9) can be rewritten as $\gamma _{\bar x,{{\bf{x}}_{e,k}}}^{{t_1}} = \frac{{{\alpha _e}{\rho _s}}}{{{\alpha _c}{\rho _s} + |\lambda _{s\bar x}^{{t_1}}{|^{ - 2}}}}$ and $\gamma _{c,{{\bf{x}}_{c,k}}}^{{t_1}} = {{\alpha _c}{\rho _s}}{{|\lambda _{sc}^{{t_1}}{|^{ 2}}}}$. Based on (16), the outage probability can be calculated as
\begin{align}
P_{{{\bf{x}}_{c,k}}}^{out} = \left\{ {\begin{array}{*{20}{l}}
{1 - \exp ( - {\xi _5}/\varOmega _{sc}^{{t_1}}),}&{{{\alpha _e}/{\alpha _c} > {\varphi _{{{\bf{x}}_{e,k}}}}}}\\
{1,}&{{{\alpha _e}/{\alpha _c} \le {\varphi _{{{\bf{x}}_{e,k}}}}}}
\end{array}} \right.
\end{align}
where ${\xi _5} = \max \Big\{ \frac{{{\varphi _{{{\bf{x}}_{e,k}}}}}}{{({\alpha _e} - {\varphi _{{{\bf{x}}_{e,k}}}}{\alpha _c}){\rho _s}}},\frac{{{\varphi _{{{\bf{x}}_{c,k}}}}}}{{{\alpha _c}{\rho _s}}}\Big\}$.

Similarly, the SINRs in (12) and (13) become $\gamma _{c,{{{\bf{\bar x}}}_{c,k}}}^{{t_2}} = {\rho _s}|\lambda _{sc}^{{t_2}}{|^2}$ and $\gamma _{e,{{\bf{x}}_{e,k}}}^{{t_2}} = {{\rho _r}}{{|\lambda _{re}^{{t_2}}{|^{2}}}}$ for this case, thus the outage probabilities $P_{{{\bf{x}}_{e,k}}}^{out}$ and $P_{{{{\bf{\bar x}}}_{c,k}}}^{out}$ can be respectively written as
\begin{align}
P_{{{\bf{x}}_{e,k}}}^{out} &={\kern -3pt} \left\{ {\begin{array}{*{20}{l}}
\begin{array}{l}
{\kern -10pt} 1 - \exp \Big(- \frac{{{\varphi _{{{\bf{x}}_{e,k}}}}}}{{{\rho _r}\varOmega _{re}^{{t_2}}}}\Big)\\
 {\kern -5.5pt} \times \exp \Big( - \frac{{{\varphi _{{{\bf{x}}_{e,k}}}}}}{{({\alpha _e} - {\varphi _{{{\bf{x}}_{e,k}}}}{\alpha _c}){\rho _s}\varOmega _{sr}^{{t_1}}}}\Big),
\end{array}&{{\kern -9pt}{\alpha _e}/{\alpha _c} > {\varphi _{{{\bf{x}}_{e,k}}}}}\\
{{\kern -4.5pt}1,}&{{\kern -9pt}{\alpha _e}/{\alpha _c} \le {\varphi _{{{\bf{x}}_{e,k}}}}}
\end{array}} \right.\\
P_{{{{\bf{\bar x}}}_{c,k}}}^{out} &= 1 - \exp \Big(-\frac{{{\varphi _{{{{\bf{\bar x}}}_{c,k}}}}}}{{{\rho _s}\varOmega _{sc}^{{t_2}}}}\Big)
\end{align}
\subsection{Outage Sum Rate}
The outage sum rate is one of the important metrics for characterizing the system spectral efficiency. Without loss of generality, we assume that $NM$ DD domain symbols ${{\bf{x}}_{c,k}}$, $1 \le k \le NM$, have the same target date rate, i.e., ${R_{{{\bf{x}}_{c,k}}}}$, as do ${{\bf{x}}_{e,k}}$ and ${{\bf{\bar x}}_{c,k}}$. Therefore, the the normalized outage sum rate for the proposed scheme can be defined as
\begin{align}
SR = &\frac{1}{{2NM}}\sum\nolimits_{k = 1}^{NM} \big\{ (1 - P_{{{\bf{x}}_{c,k}}}^{out}){R_{{{\bf{x}}_{c,k}}}}
\\ \nonumber
&+ (1 - P_{{{\bf{x}}_{e,k}}}^{out}){R_{{{\bf{x}}_{e,k}}}} + (1 - P_{{{{\bf{\bar x}}}_{c,k}}}^{out}){R_{{{{\bf{\bar x}}}_{c,k}}}}\big\}
\end{align}
\section{Numerical Results}
This section presents extensive Monte Carlo simulations to evaluate the outage probability and outage sum rate of the proposed OTFS-NOMA scheme for CDRT. Specifically, we consider an OTFS-NOMA system with a carrier frequency of 4 GHz and a subchannel spacing of 3.75 KHz. The largest Doppler shift corresponding to the maximal speed 126.56 Km/h is ${v_{\max }} = 468.74$ Hz. Moreover, we set $\varOmega _{sc}^{{t_1}} = \varOmega _{sc}^{{t_2}} = \varOmega _{re}^{{t_2}} = 1$, $\varOmega _{sr}^{{t_1}} = 0.5$, and ${P_r}/{P_s} = 0.5$. The power allocation coefficients are assumed to ${\alpha _e} = 0.9$ and ${\alpha _c} = 0.1$.

Fig. 2 compares the outage probabilities for the proposed scheme (i.e., Prop.) and the OTFS-based OMA (OTFS-OMA) benchmark in high-mobility scenarios, where the OTFS-OMA employ four phases to complete the transmission of the signals ${{\bf{x}}_{c,k}}$, ${{\bf{\bar x}}_{c,k}}$, and ${{\bf{x}}_{e,k}}$, $1 \le k \le MN$, in the proposed scheme.
The figure shows that the analytic expressions are consistent with the corresponding simulation results.
Under different parameter settings, the proposed scheme achieves better outage performance for $U_c$ and $U_e$ in $t_2$ (i.e., ${{\bf{\bar x}}_{c,k}}$ and ${{\bf{x}}_{e,k}}$) than the OTFS-OMA scheme.
 This is because the OTFS-OMA scheme involves more phases, which leads to higher decoding requirements for ${{\bf{\bar x}}_{c,k}}$ and ${{\bf{x}}_{e,k}}$.
Additionally, since the outage probability for ${{\bf{x}}_{c,k}}$ (i.e., $P_{{{\bf{x}}_{c,k}}}^{out}$) depends on the corresponding channel quality and the target data rates of ${{\bf{x}}_{e,k}}$ and ${{\bf{x}}_{c,k}}$, the proposed scheme can obtain lower $P_{{{\bf{x}}_{c,k}}}^{out}$ than OTFS-OMA when ${R_{{{\bf{x}}_{c,k}}}}$=1.8 BPCU and ${R_{{{\bf{x}}_{e,k}}}}={R_{{{\bf{\bar x}}_{c,k}}}}=1$ BPCU, while
the opposite result occurs if ${R_{{{\bf{x}}_{c,k}}}}={R_{{{\bf{x}}_{e,k}}}}=0.6$ BPCU and ${R_{{{\bf{\bar x}}_{c,k}}}}$=0.3 BPCU.
Therefore, the outage performance associated with ${{\bf{x}}_{c,k}}$ for the proposed scheme can be guaranteed by adjusting the target data rates. Fig. 3 depicts the curves of the normalized outage sum rates for the proposed scheme, OTFS-OMA and OTFS-nCDRT, where the benchmark OTFS-nCDRT utilizes the NOMA-nCDRT in \cite{R2} and OTFS modulation to transmit ${{\bf{x}}_{c,k}}$ and ${{\bf{x}}_{e,k}}$ in two phases.
The figure indicates that the proposed OTFS-NOMA scheme acquires the best outage sum rate among all schemes, which means that the proposed scheme has the advantage of high spectral efficiency due to the non-orthogonal multiplexing of multi-user power domain resources.
Moreover, the outage sum rate is dominated by the target data rates and the outage probabilities, and the outage probabilities gradually decrease as the SNR increases.Therefore, the outage sum rate possesses a ceiling equaling to the sum of all the target data rates for high SNR.
\vspace*{-3pt}
\section{Conclusion}
This paper has proposed a OTFS-NOMA scheme for the CDRT system with high mobility users. The closed-form expressions of the outage probability and outage sum rate for the proposed scheme have been derived to characterize the system performance. Numerical results have verified the theoretical
analysis and demonstrated that the proposed scheme has the performance superiority over the OTFS-OMA benchmark in terms of the outage probability for the far user. Besides, the proposed scheme can achieve the highest outage sum rate compared with the OTFS-NOMA and OTFS-nCDRT schemes.
\begin{figure}[t]
\centering
\subfigbottomskip=-0.1pt
	\subfigcapskip=-5pt
\begin{minipage}[t]{0.499\textwidth}
\centering
{\subfigure[]{
         \includegraphics[width=0.48\textwidth]{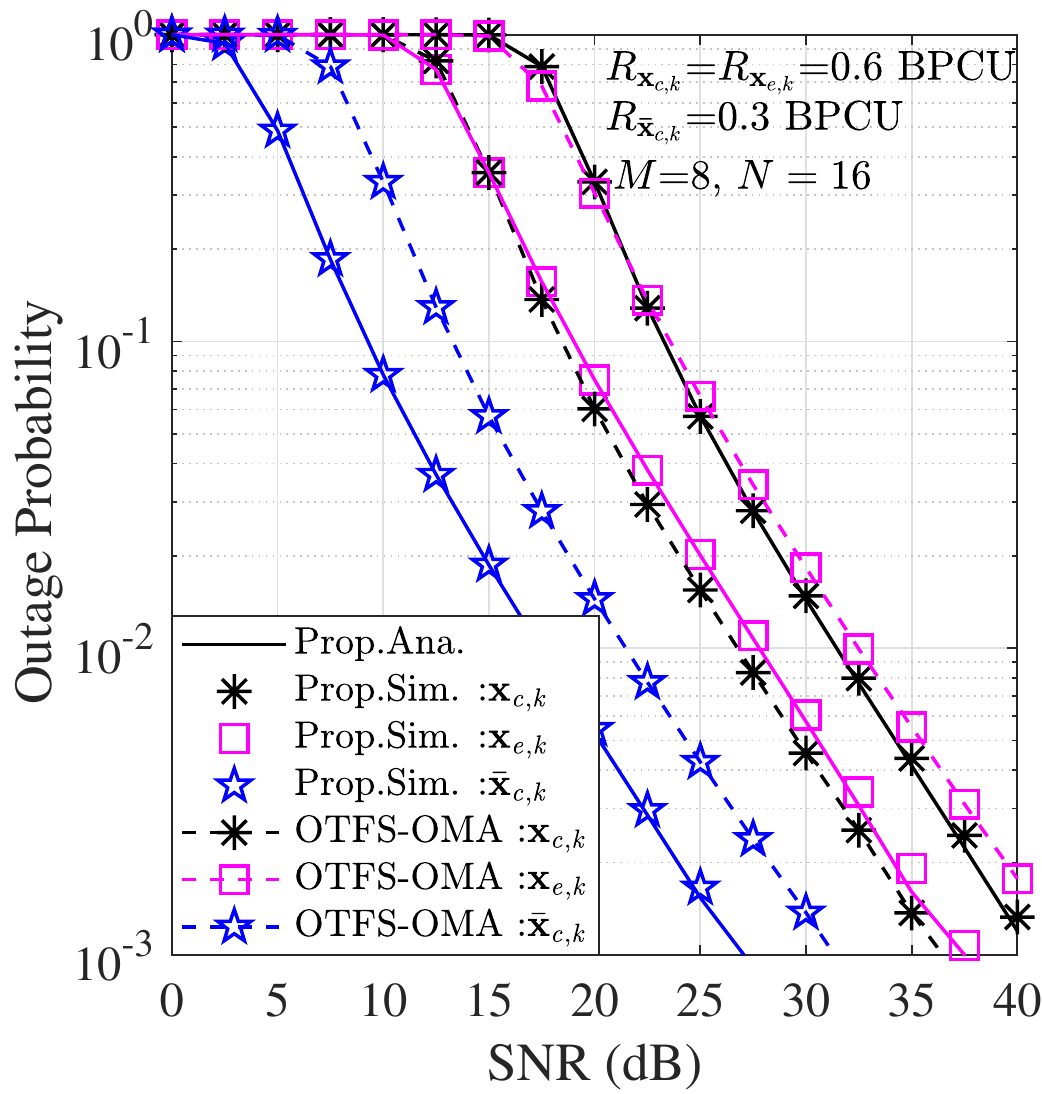}}
\subfigure[]{
         \includegraphics[width=0.48\textwidth]{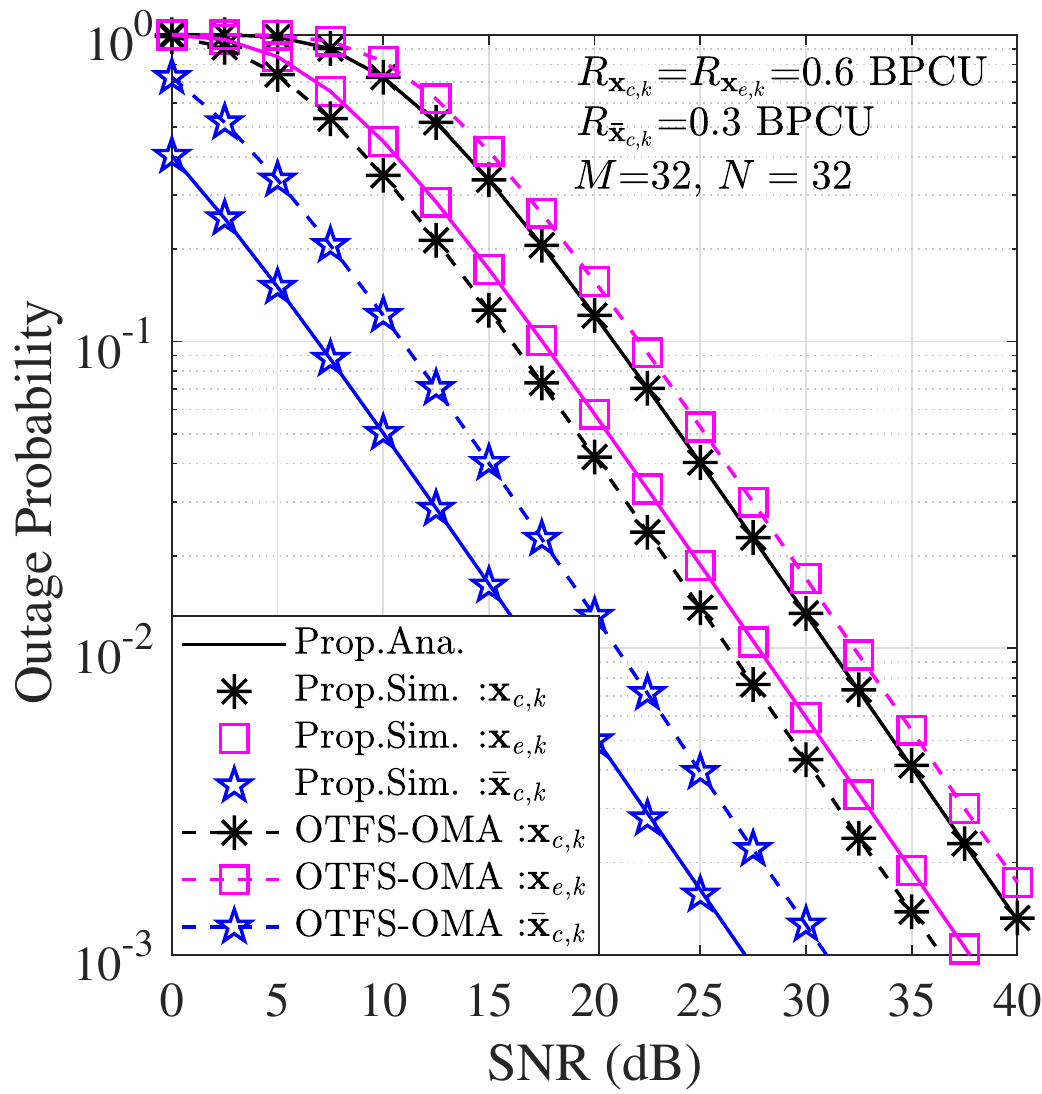}}}
{\subfigure[]{
         \includegraphics[width=0.48\textwidth]{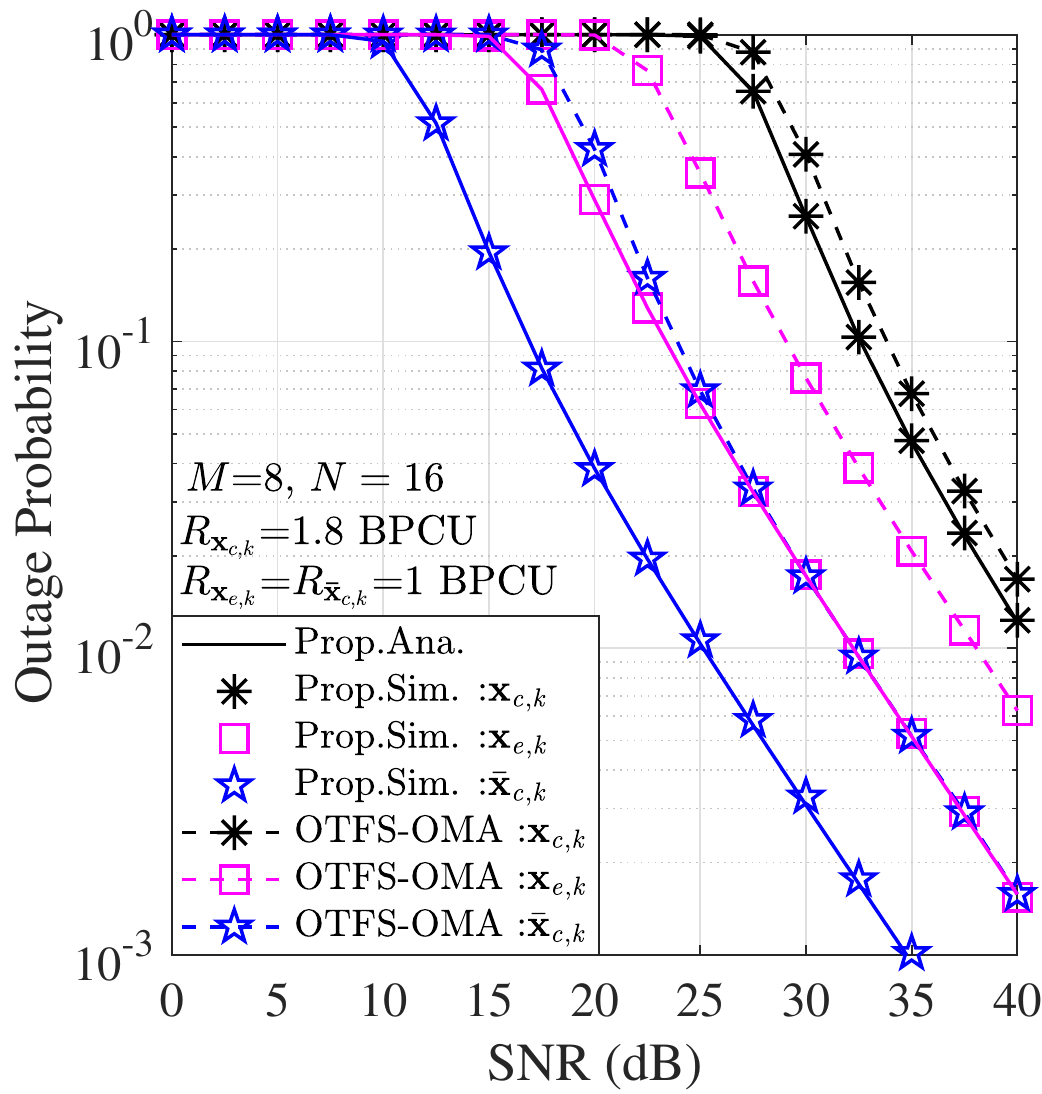}}
\subfigure[]{
    \includegraphics[width=0.48\textwidth]{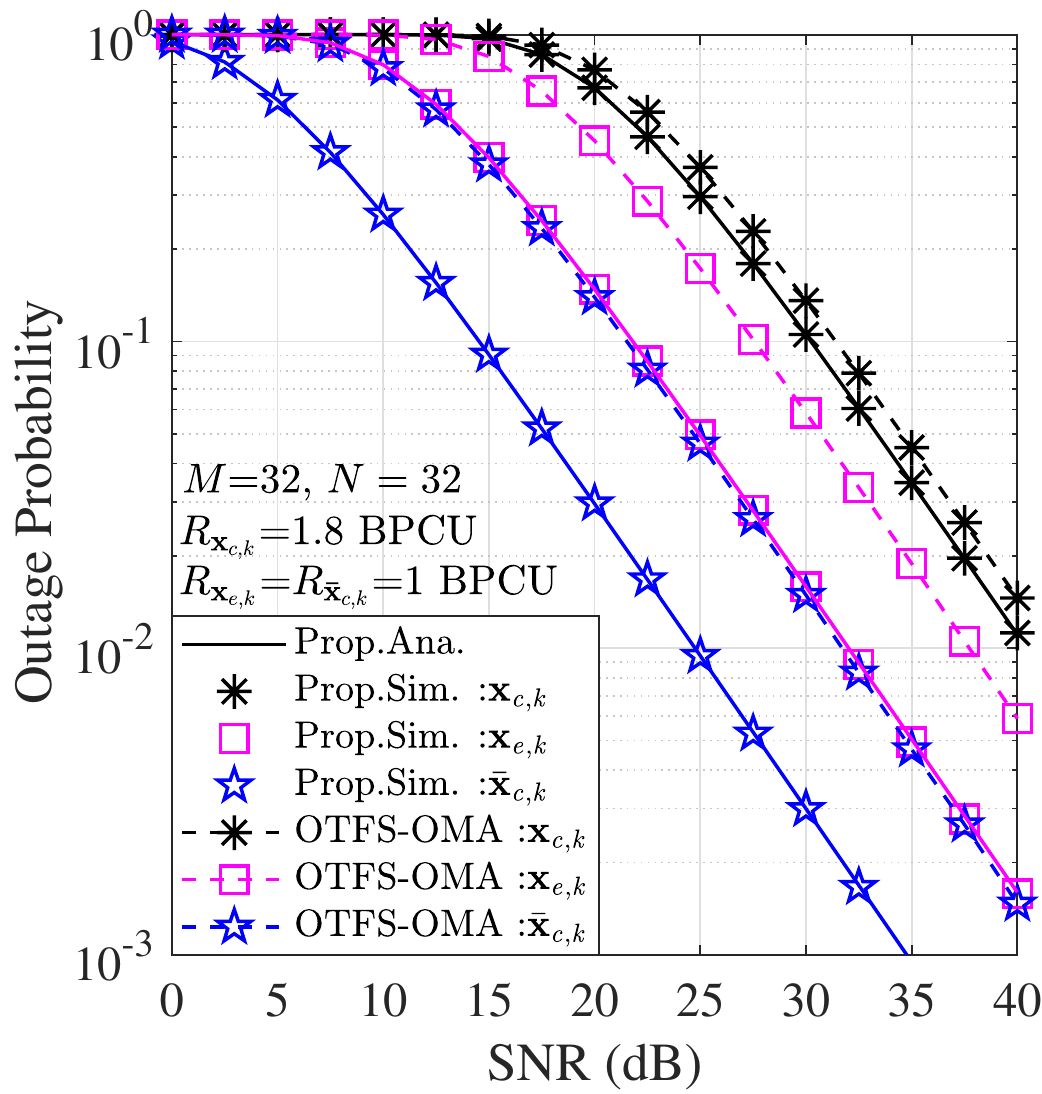}}}
\end{minipage}
\vspace{-0.1in}
\caption{Comparison of outage probability between the proposed scheme and OTFS-OMA. For (a) and (c) General Case with ${P_{xy}^i}$=3, $k_{xy}^{i,\omega}$=[0 1 2], $l_{xy}^{i,\omega }$=[0 2 3], and for (b) and (d) Special Case with ${P_{xy}^i}$=1, $k_{xy}^{i,\omega}$=[1], $l_{xy}^{i,\omega }$=[1].}
\vspace{-0.2in}
\end{figure}
\begin{figure}[t]
\centering
\subfigbottomskip=-0.1pt
	\subfigcapskip=-5pt
\begin{minipage}[t]{0.499\textwidth}
\centering
{\subfigure[]{
         \includegraphics[width=0.48\textwidth]{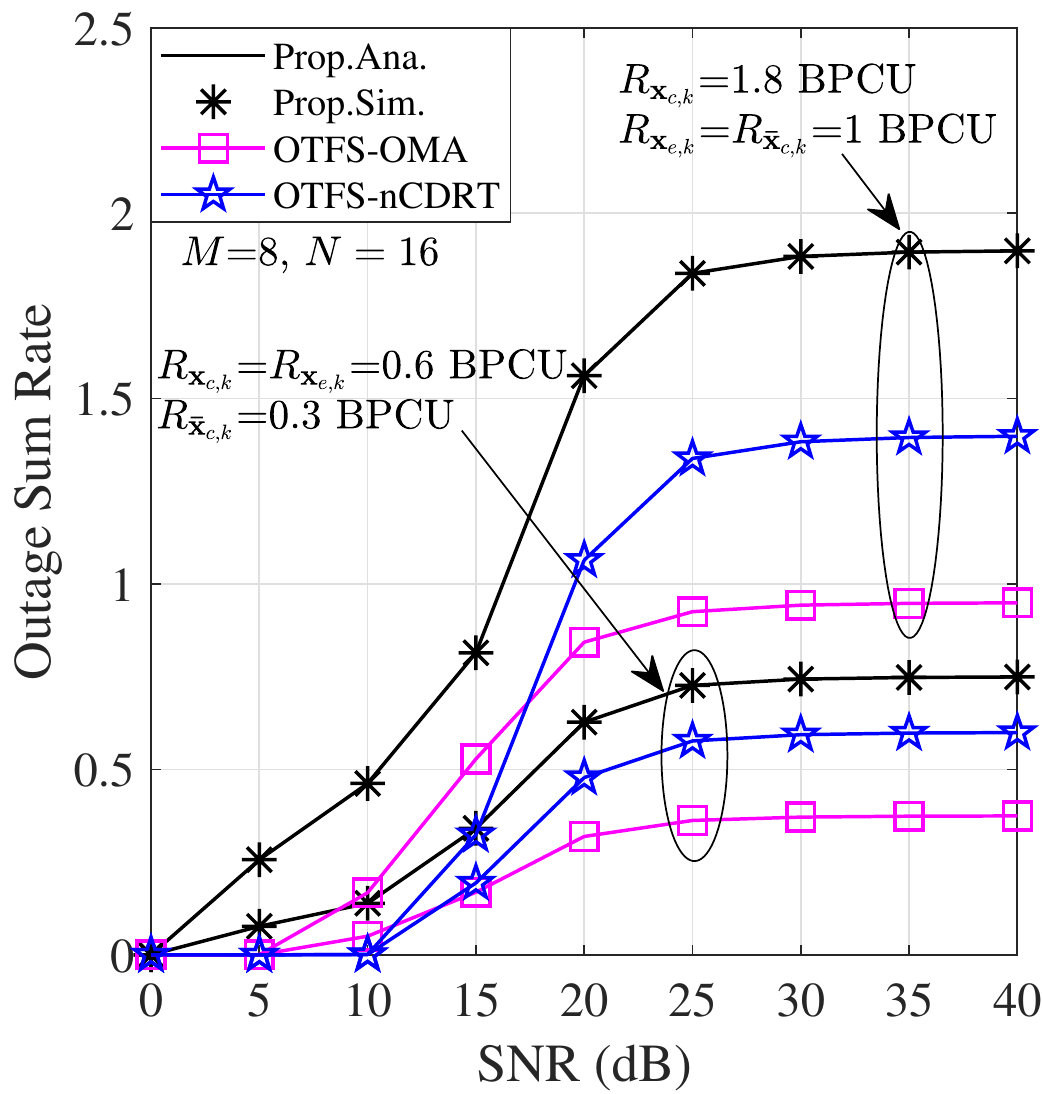}}
\subfigure[]{
         \includegraphics[width=0.48\textwidth]{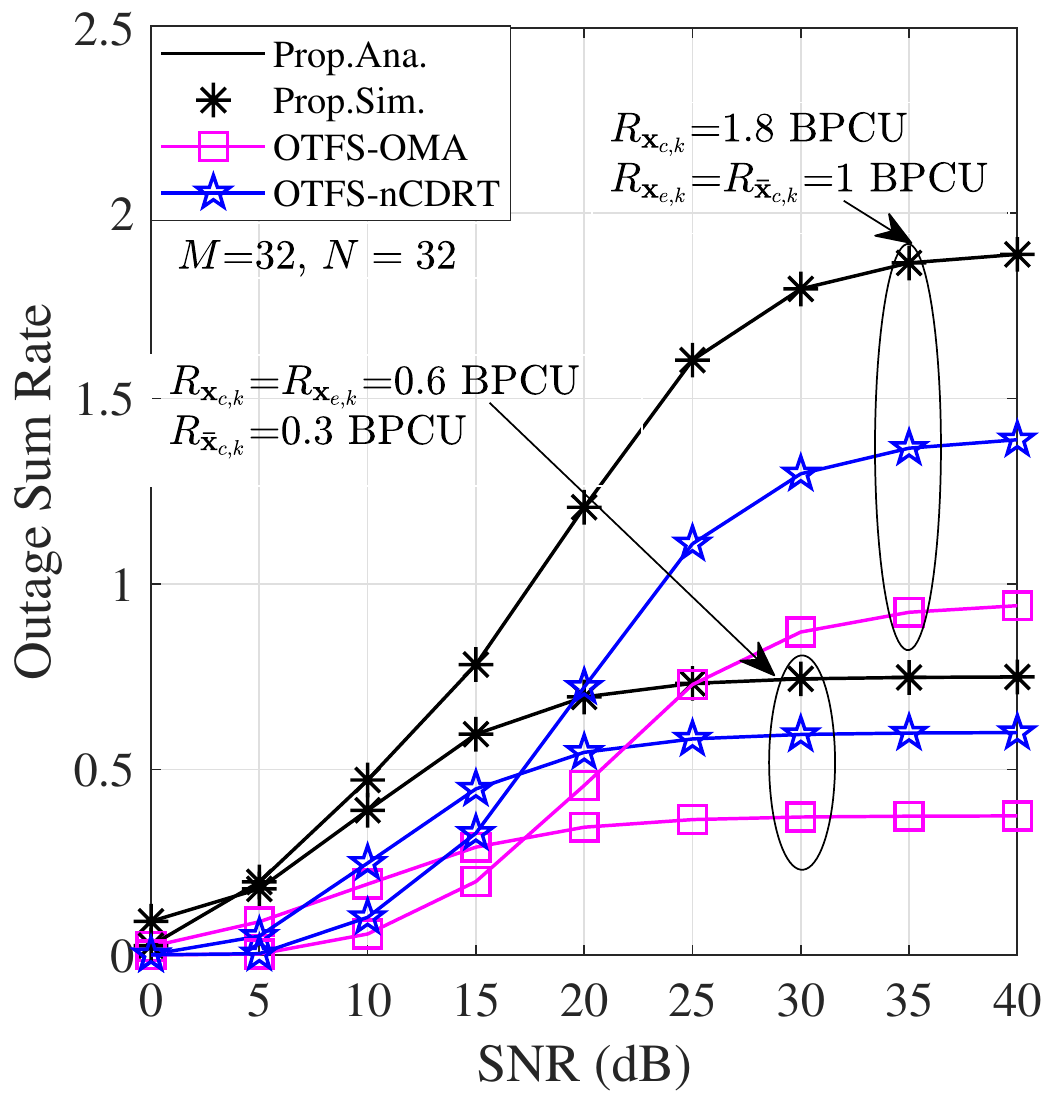}}}
\end{minipage}
\vspace{-0.1in}
\caption{ Outage sum rate versus SNR: (a) General Case with ${P_{xy}^i}$=3, $k_{xy}^{i,\omega}$=[0 1 2], $l_{xy}^{i,\omega }$=[0 2 3]; and (b) Special Case with ${P_{xy}^i}$=1, $k_{xy}^{i,\omega}$=[1], $l_{xy}^{i,\omega }$=[1].}
\vspace{-0.1in}
\end{figure}
\vspace*{-2pt}

\ifCLASSOPTIONcaptionsoff
  \newpage
\fi



%
\appendices

\setlength{\parskip}{0.1\baselineskip}

%








\end{document}